\documentclass[english]{achemso}
\usepackage{lmodern}
\usepackage[LGR,T1]{fontenc}
\usepackage[latin9]{inputenc}
\usepackage{textcomp}
\usepackage{amssymb}
\usepackage{graphicx}

\makeatletter

\title{Arrested coalescence of viscoelastic droplets: Ellipsoid shape effects
and restructuring}
\author{Chen Hao}
\affiliation{School of Chemical Engineering, UNSW Sydney, NSW, Australia}
\author{Zhaoyu Xie}
\affiliation{Dept. Physics and Astronomy, Tufts University, Boston, MA, USA}
\author{Timothy J. Atherton}
\affiliation{Dept. Physics and Astronomy, Tufts University, Boston, MA, USA}
\author{Patrick T. Spicer}
\affiliation{School of Chemical Engineering, UNSW Sydney, NSW, Australia}
\email{p.spicer@unsw.edu.au}
\DeclareRobustCommand{\greektext}{%
  \fontencoding{LGR}\selectfont\def\encodingdefault{LGR}}
\DeclareRobustCommand{\textgreek}[1]{\leavevmode{\greektext #1}}
\ProvideTextCommand{\~}{LGR}[1]{\char126#1}

\makeatother

\usepackage{babel}
\begin{document}
\begin{abstract}
The stable configurations formed by two viscoelastic, ellipsoid-shaped
droplets during their arrested coalescence has been investigated using
micromanipulation experiments. Ellipsoidal droplets are produced by
millifluidic emulsification of petrolatum into a yield stress fluid
that preserves their elongated shape. The liquid meniscus between
droplets can transmit stress and instigate movement of the droplets,
from their initial relative position, in order to minimize doublet
surface energy. The action of the liquid meniscus causes the ellipsoidal
droplets to undergo rolling and restructuring events because of their
unique ellipsoid shape and associated variation in surface curvature.
The final configuration of the droplets is shown to be controlled
by the balance between interfacial Laplace pressure and internal elasticity,
as well as a constraint force that resists complete minimization of
surface energy. Geometric and surface energy calculations are used
to map the possible and most likely configurations of the droplet
pairs. Experimental deviations from the calculations indicate the
magnitude and potential origin of the constraint force resisting full
equilibration. Droplet aspect ratio and elasticity are both shown
to influence the degree of restructuring and stability of the droplets
at energy extrema. Higher aspect ratios drive greater restructuring
and better agreement with final doublet configurations predicted by
energy minimization. Lower elasticity droplets undergo secondary deformations
at high aspect ratios, further broadening the space of possible morphologies. 
\end{abstract}

\section*{Introduction}

Emulsions are broadly used in food \cite{Friberg2004}, pharmaceutical
\cite{gupta2016nanoemulsions}, polymeric \cite{Vandebril2010}, petroleum
\cite{zylyftari2015modeling}, biomimetic \cite{Claessens2006,pontani2012biomimetic},
bijel \cite{Mohraz:2016hd}, and even explosive \cite{Bdzil2007}
materials. Coalescence, when emulsion droplets recombine into a larger
fluid volume, is a critical dynamic in commercial products that can
be harmful, when product instabilities occur or dispersion quality
is lost, or beneficial, when a structure of linked droplets imparts
rheology modification and desirable texture to a fluid like, for example,
whipped cream. 

Arrested coalescence occurs when coalescence begins minimizing droplet
surface energy but is halted by a physical resistance like interfacial
or internal droplet elasticity \cite{Studart2009,Pawar2011,Pawar2012}.
Arrested coalescence is commonly observed in dairy products, where
a network of solid milk fat crystals forms inside of oil droplets
\cite{Boode:1993vp}. The poroelasticity of the droplet allows coalescence
to begin because of the liquid phase permeating the solid crystal
skeleton, but is then halted when the crystal structure can not be
deformed enough for full coalescence \cite{Pawar2011,Pawar2012}.
Arrested coalescence is important to the study of food products, as
it is a key phenomenon used to build structure, texture, and aesthetic
perception of most dairy products \cite{Fredrick2010}, but recent
work has shown applications of arrested droplets in other materials
as well. 

Colloidal materials are often developed under conditions when arrested
coalescence is significant, as when partially molten metal particles
assemble into clusters \cite{VanderKooij2015} or when evaporating
films of solvent squeeze suspended particles or droplets \cite{Manoharan2003,Sacanna2010,Sacanna2011,Nabavi2017}.
The final structure and performance of such systems is highly dependent
on the packing and structure of constituent spherical colloids, and
it is important to know the effects of dynamic capillary processes
in these systems on the evolution of structure as assembly proceeds.
For example, it was shown recently that spherical viscoelastic droplets
can significantly reorient as a result of liquid capillary effects
during arrested coalescence, spontaneously driving the droplets into
more compact packing arrangements even though initially added at less-optimal
positions \cite{Dahiya:2017gk}. However, meniscus effects also play
a significant role in shape changes by aggregates of non-spherical
solid colloids, enhancing self-assembly \cite{Basavaraj2006,Madivala2009,Yunker2011,Phillips2014},
sintering of metal structures \cite{Nabavi2017}, additive manufacturing
\cite{Prileszky:2016gc}, and aggregation of non-spherical fat solids
\cite{Kim2013}. 

Droplets are typically spherical because interfacial tension acts
to minimize surface energy. In some viscoelastic droplets, however,
the internal rheology of the droplets can preserve non-spherical shapes
\cite{Caggioni2014,Caggioni2015,bayles2018model}, and is used to
enhance deposition onto biological surfaces \cite{spicer17shape,spicer17uspto}
and food emulsion rheology \cite{Thivilliers2007,Thivilliers:2008vk,ThivilliersArvis:2010vi}.
There is then also a need to understand arrested coalescence and its
shape-change dynamics for non-spherical droplets. We are interested
in these dynamics for two reasons. One is to understand how non-spherical
droplets assemble in systems whose rheology, texture, and quality
are determined by droplet microstructure. A second is to map the motion
of non-spherical shapes driven by the competing effects of interfacial
liquid meniscus motion and droplet mechanical properties. Dynamic
shape change using physical mechanisms, like geometry and interfacial
driving forces \cite{J.-P.Peraud2014}, will be a critical aspect
of future directed and active material assembly efforts at the nanoscale,
where Brownian motion is significant \cite{Cho2010}, and at the microscale,
where thermal motion no longer dominates \cite{Py:2007ts,Leong2007}. 

We study here the binary arrested coalescence of monodisperse ellipsoidal
droplets with controlled aspect ratios produced in a microfluidic
device. Because of the importance of motion to shape-change, and of
final state to self-assembly, we examine both aspects here. We first
explore the dynamics of an ellipsoidal droplet being pulled around
another by a fluid meniscus as a means of understanding possible shape
change and self-assembly mechanisms. Many more configurations are
possible for arrested pairs of ellipsoidal droplets than for spherical
droplets, and we use the results of an energy minimization model to
map the full extent of possible behavior while comparing with our
experimental findings. 

\begin{figure}
\includegraphics[scale=0.4]{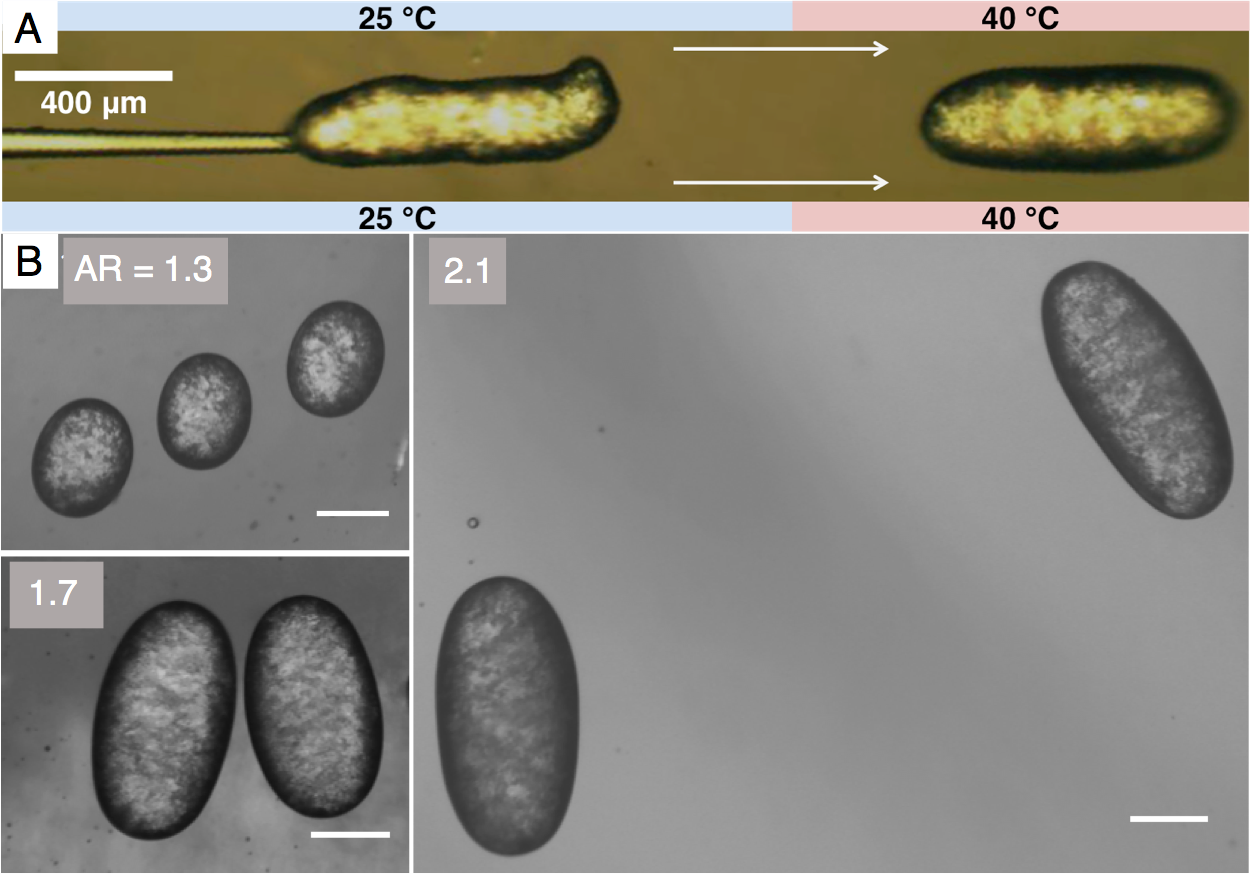}

\caption{\label{fig:fluidicAR}A) Microscopic images within the millifluidic
setup used to produce ellipsoidal droplets. An external yield stress
fluid is used as a liquid mold to retain the high droplet aspect ratio
during flow and heating. Melting after droplet production ensures
a uniformly ellipsoidal shape is produced, even with non-uniform starting
shapes like the one on the left. B) Images of three different aspect
ratio droplets produced by the millifluidic setup. Scale bar is $200\:\mu m$.}
\end{figure}

\section*{Experimental}

\subsection*{Methods and Materials }

Oil-in-water emulsions were prepared by combination of an oil and
aqueous phase during microchannel flow. The system has been characterized
in a number of past works \cite{Pawar2012,Caggioni2014,Caggioni2015,dahiya2016arrested,Dahiya:2017gk}
and is a reproducible model of more complex systems like milk fat
\cite{thiel2016coalescence}. The aqueous phase was a dispersion of
neutralized 0.15\% w/w Carbopol 846. The oil phase was a mixture of
petrolatum containing 50\% wax solids (Unilever) and varying levels
of hexadecane (99\% Sigma Aldrich) to form mixtures containing either
30\% or 40\% w/w wax solids. The two components were heated to 80
\textdegree C until the petrolatum melted and fully mixed with the
hexadecane. The mixture then crystallized as a dispersion of solid
crystals inside a liquid oil phase \cite{Pawar2012}.

\subsection*{Millifluidics }

Droplets were generated in a two-part co-flow millifluidic device
made from a round 1.1 mm ID glass capillary (Vitrocom), with a 90
\textgreek{m}m ID tip created using a Model P-97 Micropipette Puller
(Sutter Instruments), inserted into a 1.1 mm ID square capillary (Vitrocom).
Oil phase was pumped through the inner, round capillary while the
continuous phase flows through the outer square tube \cite{Utada2005}.
Droplets with varying size and aspect ratio were produced by adjusting
the dispersed phase flow rate between 0.08-1.2 mL/min while keeping
the continuous phase at 0.08 mL/min. 

As seen in Figure \ref{fig:fluidicAR}A, droplets with irregular shapes
were first produced as a result of the yield stress of the droplet
phase ($\sigma_{y}=1-500\:Pa$) \cite{Caggioni2014,Caggioni2015}.
Relatively uniform ellipsoidal shapes were then produced by immediately
passing the dispersion through a section of the channel held at 40
\textcelsius{} by an external heating source \cite{Prileszky:2016kb,Caggioni2018}.
The internal structure of the droplet partially melts during this
step, reducing its yield stress and reshaping the droplet to a uniform
ellipsoid with a maximum curvature set by the equality of the droplet
and continuous phase yield stress values \cite{Caggioni2014,Caggioni2015,bayles2018model}.
The yield stress of the $0.15\%$ w/w aqueous Carbol 846 continuous
phase ($\sigma_{y}=4.25\:Pa$) created a ``liquid mold'' environment
that preserved the deformed droplet shape until it could cool and
regain its own yield stress \cite{Pairam2013,Burke2015}. Image analysis
of calibrated experimental micrographs was carried out using ImageJ
software \cite{Schneider2012} to quantify droplet size, aspect ratio,
and long axis orientation. Doublets are studied only when they are
level in the image plane to avoid errors from out-of-plane movements,
consistent with our theoretical descriptions.

Figure \ref{fig:fluidicAR}B shows three examples of different aspect
ratio ellipsoids produced using this method. Coalescence studies were
carried out by carefully transferring droplets to a volume of aqueous
0.3\% w/w microfibrous cellulose yield stress fluid \cite{Emady2013}.
Micromanipulation was performed on droplets, inside a small liquid
sample placed on glass slides, using a 3-axis system (Narishige International)
mounted on an inverted microscope (Motic AE31). Droplets were grasped,
using suction applied by adjusting the height of a small fluid reservoir
connected to a microcapillary in the sample volume, and then the droplets
were pushed together by slow manipulation to initiate coalescence
\cite{Pawar2011}.

\section*{Results and Discussion}

Arrested coalescence can occur when the elasticity of an internal
solid network in two droplets is sufficient to balance the interfacial
Laplace pressure driving the coalescence process \cite{Pawar2012}.
As a result, the balance can be altered by, for example, increasing
the internal solids concentration of the droplets in order to increase
the droplet elasticity. When the symmetry of the system is broken,
for example when a third droplet is added to an existing arrested
pair, restructuring can occur when the meniscus pulls the droplets
in a direction that minimizes the assembly's surface energy \cite{Dahiya:2017gk}.
We consider the potential for such effects on ellipsoidal droplets
here by studying the simplistic case of an ellipse rolling around
a second one.

\subsection*{Rolling ellipsoid model}

If the experimental droplets behave as hard solids when moved around
one another, a simple rolling model can be used to study some aspects
of droplet motion. It also provides a basis for comparison with experimental
results using droplets that can undergo some degree of deformation
as a result of their internal elastic network \cite{Pawar2012}. The
two-dimensional path of the center of an ellipse rolling around another,
congruent, ellipse is described in Cartesian coordinates by the curve
\cite{abbena2017modern}:

\begin{equation}
\left(x^{2}+y^{2}\right)^{2}=4a^{2}x^{2}+4b^{2}y^{2}\label{eq:ellipse}
\end{equation}

where $a$ and $b$ are the major and minor axes of each ellipse and
the aspect ratio is defined as:

\begin{equation}
AR\equiv\frac{a}{b}\label{eq:ar}
\end{equation}

\begin{figure*}
\includegraphics[scale=0.5]{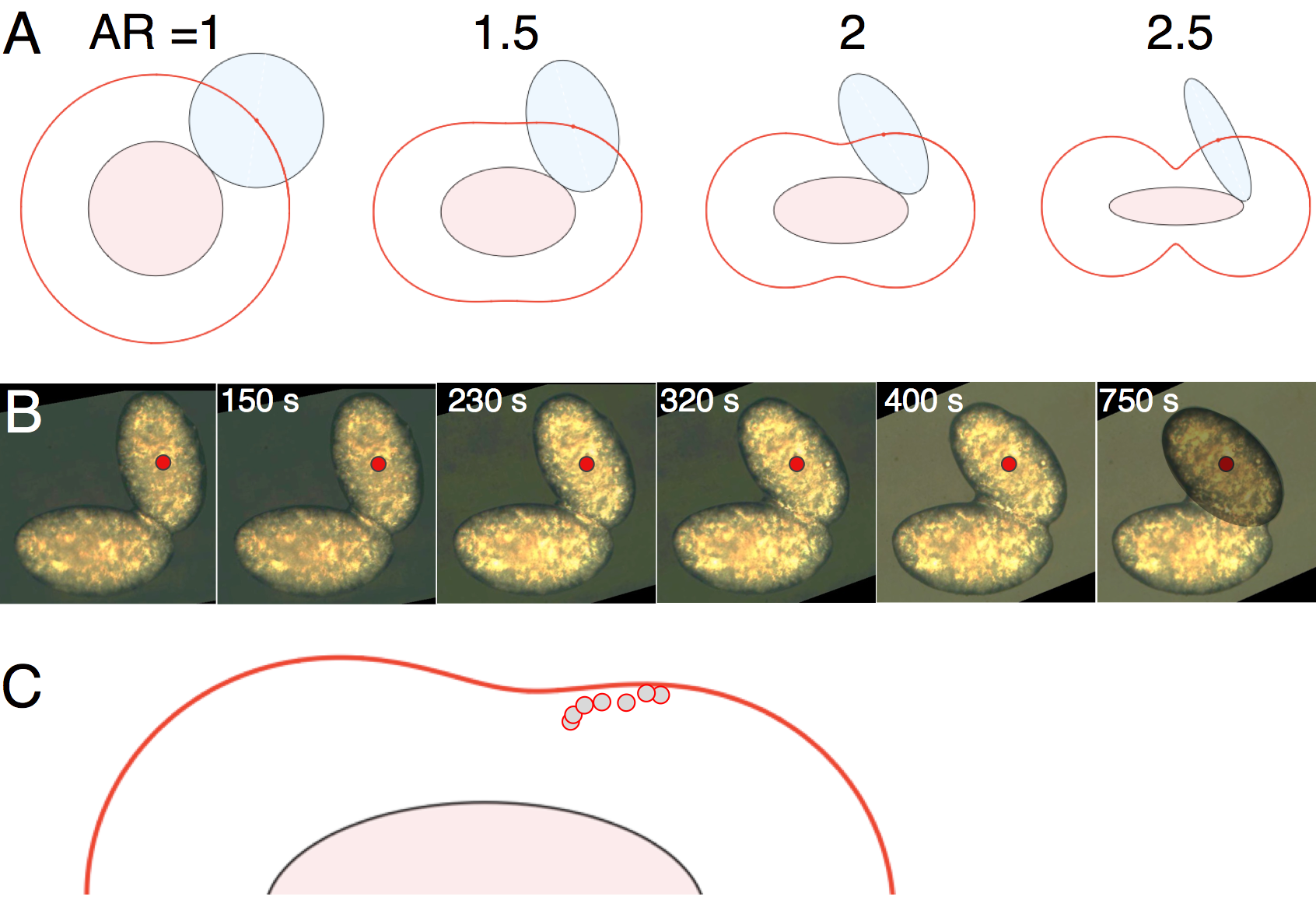}

\caption{\label{fig:Path-plot}A) Paths plotted in red for the center of mass
of various ellipsoids rolling around a second one. Increasing the
aspect ratio alters the path of the center of mass, varying the path
curvature when the two shapes are aligned edge-to-edge. B) Experimental
micrographs of two ellipsoidal droplets containing $40\%$ solids
as they undergo arrested coalescence and restructuring. C) Plotted
positions of the center of mass of the upper ellipsoid in B) as compared
to the calculated path of a system with the same dimensions using
Equation \ref{eq:ellipse}.}
\end{figure*}

Equation \ref{eq:ellipse} assumes motion of the ellipses begins with
the two shapes aligned end-to-end along their long axes and ignores
slip between the two shapes during motion. Visualizations of calculated
ellipse paths were produced using code developed by Mahieu \cite{Mahieu2014}.
Figure \ref{fig:Path-plot}A shows several plots of the calculated
trajectory, in red, of the center of mass of a blue ellipse, rolling
around the perimeter of a second, identical, red ellipse. As the aspect
ratio of the ellipses increases, the moving ellipse path transitions
from perfectly circular, for $AR=1$, to an increasingly complex path.
Regions of the path with sharp changes in curvature are evident when
the ellipses are aligned edge-to-edge, increasing in magnitude as
the ellipse aspect ratio increases. Figure \ref{fig:Path-plot}B shows
a series of images of two ellipsoidal droplets just after coalescence
has initiated. Here the images have been rotated to fix the position
of the lower droplet and enhance study of subsequent changes. The
liquid film bridging the droplets radially moves out from the point
of initial contact, reducing overall surface energy, and in the process
pulls the two droplets together. Similar behavior was seen for the
case of two spherical droplets undergoing arrested coalescence \cite{Pawar2012}.
However, here the non-spherical droplets also experience a change
in relative orientation as the interfacial force pulling them together
rolls one droplet along the edge of the other, similar to the action
of ellipsoidal gears \cite{Zar2008}. A red dot has been placed at
the center of mass of the upper ellipsoid, allowing us to track changes
in orientation as a result of the meniscus movement. The droplets
largely move as rigid bodies and never move in a way that increases
surface area, consistent with previous observations of spherical droplets
moved by meniscus effects \cite{Dahiya:2017gk}. There is also some
noticeable deformation of the droplets as the meniscus pulls them
together in the last frame of Figure \ref{fig:Path-plot}B, where
overlap with the other ellipse is apparent, also in agreement with
the behavior of spherical droplets \cite{Pawar2012}. We highlight
this effect using a drawn ellipse overlay in Figure \ref{fig:Path-plot}B
that indicates the degree of overlap and deformation of the drops
as a result of elastic deformation \cite{Pawar2012}. The last frame
in Figure \ref{fig:Path-plot}B represents the final stable position
of this doublet, even though we might naively expect the two droplets
to relax further. Some resistance to the reduction of surface energy
must be acting to halt further motion, perhaps because of local friction
or deformation of the internal elastic microstructure. Comparison
of the center of mass positions in Figure \ref{fig:Path-plot}B with
calculated paths for this doublet will allow us to quantify the magnitude
of any effects on trajectory.

In Figure \ref{fig:Path-plot}C we plot a trajectory calculated following
the method used to generate Figure \ref{fig:Path-plot}A along with
the experimentally-determined positions of the center of mass of the
upper droplet in Figure \ref{fig:Path-plot}B to assess any non-idealities.
We see good general agreement of the experimental and calculated paths,
though the real droplet is systematically closer to the second droplet
than predicted. The deviation likely results from the small strain
the droplets experience as the internal solid network deforms elastically
\cite{Pawar2012}. The last few points in Figure \ref{fig:Path-plot}C
deviate more significantly from the calculated path, indicating there
may be stronger deformation occurring late in the process and that
such deviations are time- and position-dependent. Variations in the
deformation during arrested coalescence between the two droplets will
cause variations in frictional resistance to movement, complicating
our description of the process. We do not quantify strain in these
systems, as we did for spherical droplets \cite{Pawar2012}, as the
more complex ellipsoidal shape here complicates such a description.
The two-dimensional calculations do not fully account for the three-dimensional
nature of the ellipsoids produced here, and also ignore the effects
of the liquid film between droplets. They do, however, communicate
the basic principle at work here: the movement of one ellipsoidal
droplet around another's perimeter is a strong function of the varying
curvature of the ellipsoids. Below we improve on the above description,
by simulating the three-dimensionality of the ellipsoids as well as
the effects of the liquid film that imposes the stress driving motion,
providing a more comprehensive understanding of the arrested coalescence
behavior of ellipsoidal droplets. 

\subsection*{Surface Minimization Model}

To map the energy landscape experienced by the ellipsoidal doublets
as a function of their relative orientation, we simulated the fluid-fluid
interface with \emph{Surface Evolver} \cite{brakke1992surface}\emph{.
}The droplets are modeled as one-sided level set constraints around
which the area of the fluid-fluid interface is minimized at fixed
enclosed volume as a function of the relative position of the two
droplets. In our model, we suppose the interface has uniform surface
tension. Hence the area of the interface is proportional to surface
energy and minimizing the area is equivalent to minimizing the surface
energy. In previous work, this strategy allowed us to correctly predict
the critical angle below which the menisci of two spherical droplets
overlapped with that of a third droplet being added, causing reconfiguration
of the triplet shape \cite{Dahiya:2017gk}. 

\begin{figure}[h]
\begin{centering}
\includegraphics[scale=0.4]{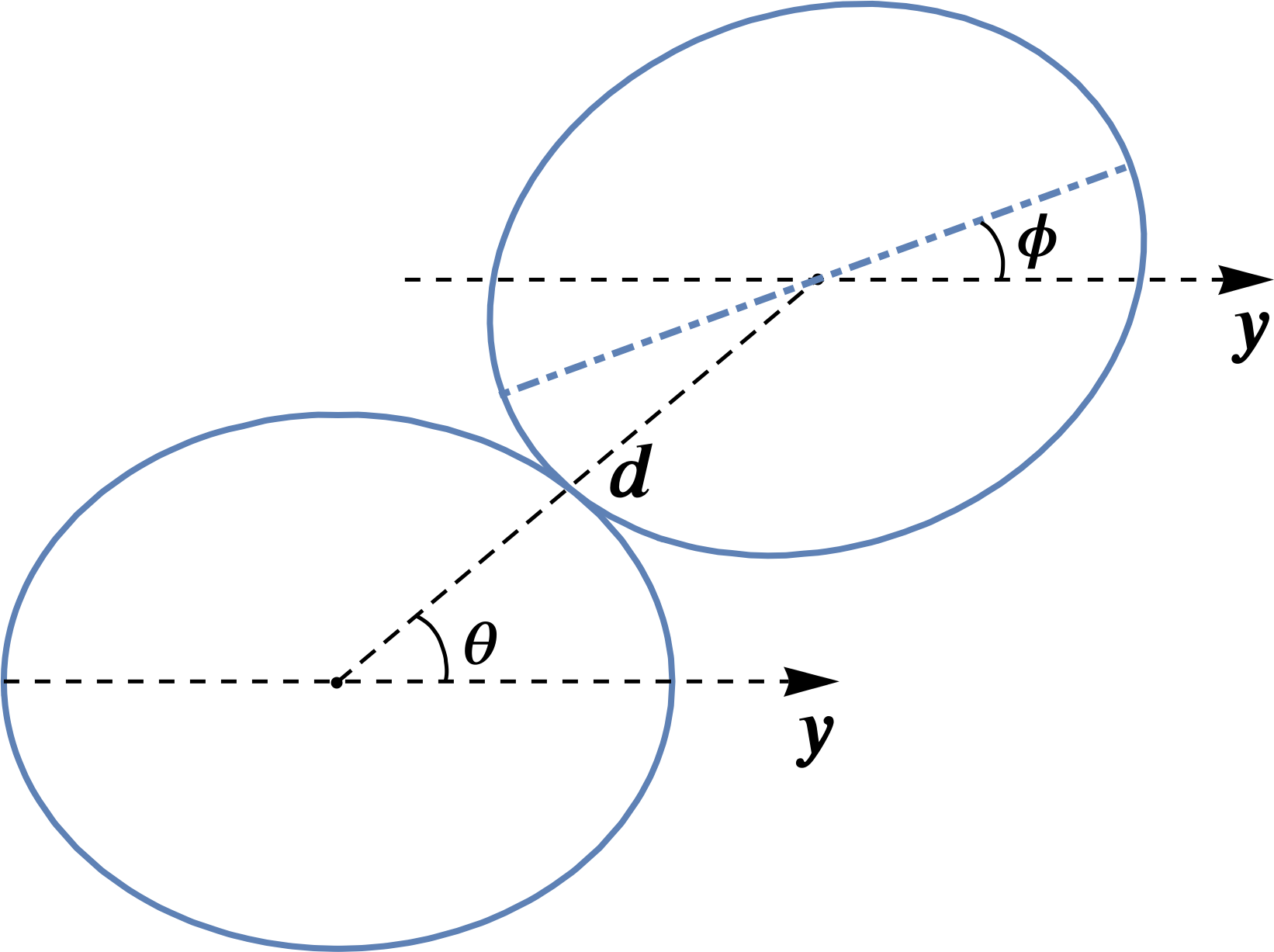}
\par\end{centering}
\caption{\label{fig:schematic}The coordinate system used to describe the relative
position of two coalescing ellipsoidal droplets.}
\end{figure}

The computational geometry is depicted in Figure \ref{fig:schematic}:
the coordinate system is oriented such that the $y$ axis is along
the major axis of one of the ellipsoids. The center of mass of the
other is at the zenithal angle $\theta$ from this axis and oriented
at an angle $\phi$ to it. Due to the symmetry of the system, the
domain of $\theta$ is from $0$ to $90^{\circ}$ and $\phi$ from
$0$ to $180^{\circ}$. For every $\theta$ and $\phi$, the distance
$d$ is calculated using the overlap algorithm proposed by Perram
and Wertheim \cite{perram1985statistical} such that the second ellipsoid
is positioned just in contact with the surface of the first ellipsoid.
The ellipsoids impose level set constraints on the fluid interface, 

\begin{equation}
\mathbf{r}\Sigma^{-1}\mathbf{r}>1
\end{equation}
\begin{equation}
(\mathbf{r}-\mathbf{r_{c}})^{T}R^{T}\Sigma^{-1}R(\mathbf{r}-\mathbf{r_{c}})>1
\end{equation}

where $\mathbf{r}=(x,y,z)$, $\mathbf{r_{c}}$ is the center of the
second ellipsoid at $(0,d\cos\theta,d\sin\theta)$, $R$ is the rotational
matrix of $\phi$ around the $x$ axis and $\Sigma$ is the diagonal
matrix with entries $(b^{2},a^{2},b^{2}$). The surface is then minimized
with a fixed volume $V=2\times\frac{4\pi}{3}ab^{2}/0.95$, slightly
greater than the volume of the two ellipsoids in order to allow the
formation of a meniscus. The volume of the enclosing fluid is clearly
a relevant quantity here, as it will set the available free liquid
to act on the droplets during their coalescence and arrest. We use
a constant value in the simulations for simplicity.

\begin{figure*}
\includegraphics[scale=0.3]{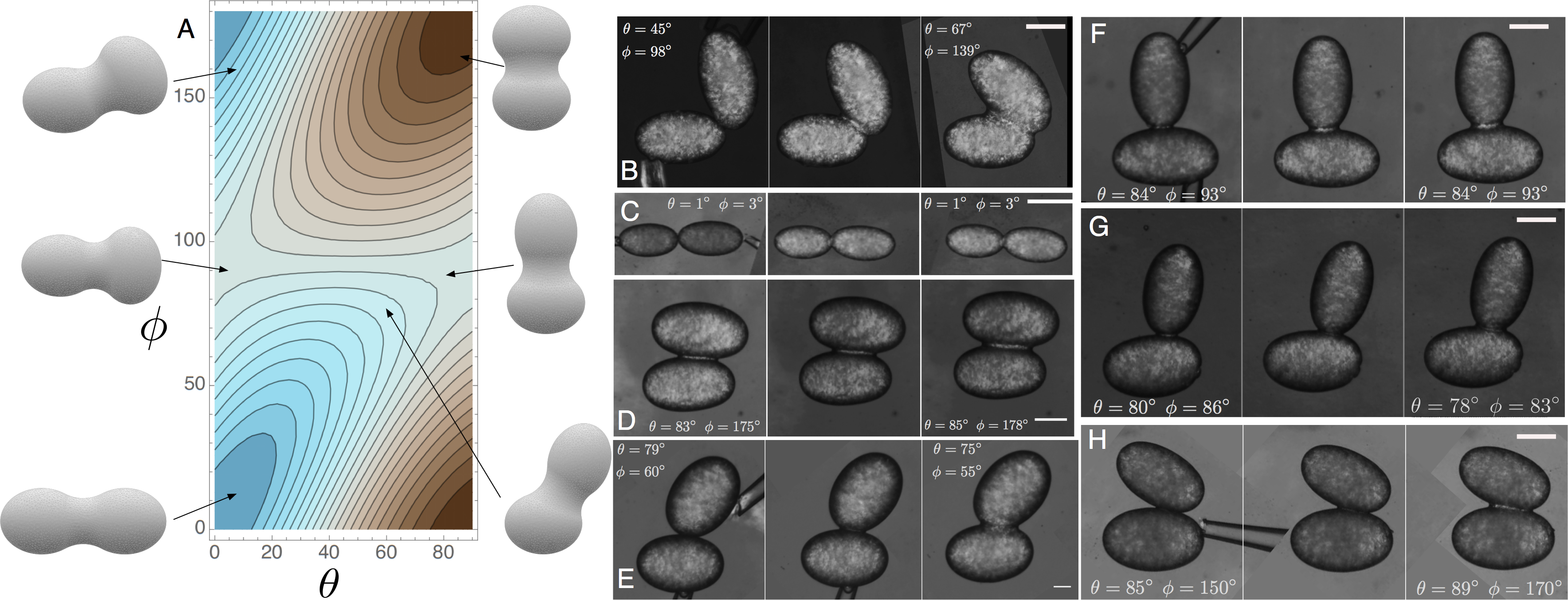}

\caption{\label{fig:Masterplot}A) A density plot of surface energy as a function
of $\theta$ and $\phi$ for ellipsoid aspect ratio $AR=1.2$. High-energy
regions are colored cyan while low-energy regions are brown, with
neutral colors indicating intermediate saddle regions. Six three-dimensional
renderings of exemplary configurations are shown for representative
energy extrema. (B-H) Stable experimental examples of some of the
simulated shapes are shown for droplets with a solids content of 40\%
and aspect ratio $AR=1.6$.}
\end{figure*}

Figure \ref{fig:Masterplot} shows a contour plot of the surface energy,
for ellipsoids of aspect ratio $AR=1.2$, calculated using \emph{Surface
Evolver} simulations for the entire range of values of angles $\theta$
and $\phi$. Three-dimensional renderings of simulated ellipsoidal
droplet pairs are also shown for representative regions to aid the
interpretation of the results and allow comparison with experiment.
Two minimum surface energy regions are visible at high $\theta$ values
in combination with either low or high $\phi$ values, consistent
with the case when two ellipsoids are joined at their long edges and
minimize their total surface area. A maximum in surface energy occurs
where the two ellipsoidal droplets are parallel and arranged end on,
at low $\theta$ values in combination with either low or high $\phi$
values, maximizing the doublet surface area. A saddle point exists
where the droplets are perpendicularly aligned in a `T' configuration.
An analogous energy diagram was computed for single nanocylinders
at a flat liquid interface, finding similar regions of stability for
similar orientations \cite{Lewandowski2006}. 

Generally, when the droplets initially make contact at arbitrary angles
we expect their orientation to evolve down the energy gradient towards
the minimum because the motion occurs in a viscous quasistatic regime,
a prediction that will be robustly tested in the subsequent section.
The droplets in the doublet are, however, also subject to forces from
the internal structure and the opposing meniscus driving forces on
the ellipsoid surface. Divergence from the behavior predicted by surface
energy alone therefore allows us to assess the relative importance
of these other terms to the final shape of the doublet. 

Figures \ref{fig:Masterplot}B-H show several examples of experimental
ellipsoidal doublets forming and restructuring, to various degrees,
before reaching their final arrested state. The droplets all have
an aspect ratio of $AR\sim1.2$ and contain $40\%$ solids by weight.
As a result, the droplet deformation is expected to be relatively
low during arrested coalescence and any observed restructuring by
the fluid meniscus will be reproducible and easily compared with the
predictions of Figure \ref{fig:Masterplot}A. Figure \ref{fig:Masterplot}A
suggests we might only find stable configurations of ellipsoidal doublets
at the two energy minima. However, Figures \ref{fig:Masterplot}B-H
show a much wider range of final doublet states can be formed, without
necessarily exhibiting the restructuring we might expect. Figures
\ref{fig:Masterplot}B and H show the largest amount of restructuring,
with $\phi$ varying by $40^{\circ}$ and $20^{\circ}$, respectively.
In Figure \ref{fig:Masterplot}B, the restructuring halts with the
two ellipsoids oriented at angles of $\theta=67{}^{\circ}$ and $\phi=139{}^{\circ}$,
close to the energy minimum in Figure \ref{fig:Masterplot}A but several
contour lines outside of it. Clearly some type of constraint force,
possibly the elastic resistance of internal structure, affects the
restructuring process and can arrest the system at a point that is
not stationary with respect to the surface tension term. In Figure
\ref{fig:Masterplot}H, an energetic minimum is attained by the pair
of ellipsoids at values of $\theta=89{}^{\circ}$ and $\phi=170{}^{\circ}$.
Other initial states lead to much less restructuring, expanding the
range of forms that can be obtained by ellipsoidal droplet arrested
coalescence. Because the ellipsoidal droplet shape is anisotropic,
this implies that the initial orientation of the droplets\textquoteright{}
major axes will have a strong effect on the subsequent restructuring
brought about by the meniscus.

Figures \ref{fig:Masterplot}C-G restructure very little from their
initial position, moving at most $5^{\circ}$ in Figure \ref{fig:Masterplot}E.
In Figure \ref{fig:Masterplot}C, two ellipsoids initiate contact
at angles of $\theta=1{}^{\circ}$ and $\phi=3{}^{\circ}$ and do
not experience any restructuring over time, despite this representing
the highest energy configuration calculated in Figure \ref{fig:Masterplot}A.
This is because the meniscus forces that normally drive restructuring
vanish at an energetic maximum, creating an unstable stationary state.
However, additional influences present, such as elastic resistance
of the internal structure, may stabilize the stationary state against
collapse. Figure \ref{fig:Masterplot}D shows the case when two ellipsoids
are initially brought together into the minimum energy configuration
at the top right of Figure \ref{fig:Masterplot}A. Here the system
restructures only very slightly. Figures \ref{fig:Masterplot} E-G
also do not restructure significantly, and they all come into contact
in various realizations of the saddle point conditions in Figure \ref{fig:Masterplot}A,
an area of intermediate stability where two ellipsoids are arrested
at, or near, right angles to one another. The system may be stabilized
at the saddle point, as in Figure \ref{fig:Masterplot}F. If the doublet
is initially configured slightly away from the saddle point, as in
Figures \ref{fig:Masterplot}E and G, the system evolves towards the
energy minimum slightly but nonetheless is arrested close to the saddle
point. Again, this may be explained by the additional constraint forces
from the internal structure. 

\begin{figure*}
\includegraphics[scale=0.8]{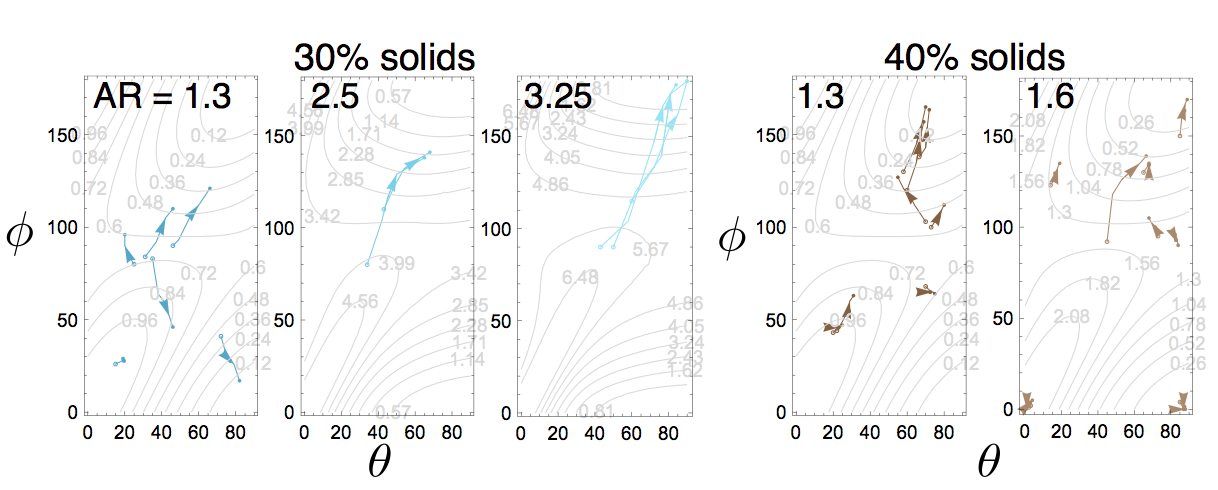}

\caption{\label{fig:trajects}Experimental pathways of ellipsoidal doublets
relaxing to their final configuration are plotted on several calculated
surface energy contour maps for different ellipsoidal aspect ratios.
Arrows indicate direction of movement on each trajectory. Increasing
the aspect ratio changes the shape of the extrema regions, as a result
of the changes in ellipsoid geometry, and also enables relaxation
to lower energy states for 30\% solids.}
\end{figure*}

Because the ellipsoid dimensions determine the amount and distribution
of surface area in the droplets, we also explore the effects of droplet
aspect ratio on our above results. Calculated energy contour plots
are shown in Figure \ref{fig:trajects} for several ellipsoidal droplet
aspect ratios, as well as two different internal droplet solids levels.
Our simulation can account for changes in solids level by varying
the volume of fluid enclosed by the simulated interface, but we use
a fixed value here because it doesn't change the shape of the energy
contour plot. The contours in each plot are labeled with surface energy
relative to the surface energy minimum. Increasing aspect ratio alters
the shape of the regions of energy minima and maxima as a result of
the geometric changes in the droplets. For example, the width of the
minimum energy region for the three aspect ratios with $30\%$ solids
in Figure \ref{fig:trajects} increases as $AR$ increases, spanning
a wider range of $\theta$ values. The effect is geometric, and is
caused by the higher ellipsoid edge length at larger $AR$ values. 

The instantaneous values of $\theta$ and $\phi$ for a number of
experimental doublets, during coalescence and restructuring, are plotted
as colored trajectory lines on top of the contour plots in Figure
\ref{fig:trajects} to show the extent of restructuring that occurs
for different starting positions and compositions. The contour plot
provides a useful map to summarize many different experiments and
compare the effects of multiple system variables. For droplets containing
$30\%$ solids, at all aspect ratios, the trajectories plotted in
Figure \ref{fig:trajects} all move in the direction of decreasing
energy. Interestingly, for $AR=1.3$, most do not reach a minimum,
consistent with the example in Figure \ref{fig:Masterplot}B, likely
because of inherent solid and fluid resistance to flow and movement.
As the aspect ratio of the droplets containing $30\%$ solids is increased
in Figure \ref{fig:trajects}, the lengths of the trajectories increase
significantly, indicating a greater degree of movement and restructuring
during arrested coalescence. Although fewer data were taken for the
larger aspect ratios, droplets with $AR=2.5$ and $AR=3.25$ reproducibly
move from an intermediate stability level, at a calculated saddle
point, entirely into the minimum energy region. Despite the simplicity
of the model, agreement between experiment and predicted final configurations
is surprisingly good when meniscus-induced restructuring is this significant.
It is reasonable to expect that longer ellipsoids will have stronger
driving forces to reduce surface energy, restructuring much more significantly
than droplets with smaller aspect ratios. The concept of a driving
force, here reduction of surface energy, and a resistance, here an
elastic or frictional force, suggests that increasing droplet solids
level and elasticity will also affect the restructuring process. 

Figure \ref{fig:trajects} plots trajectories for a number of doublet
systems containing $40\%$ solids as they restructure. For an $AR=1.3$,
the trajectory sizes are of similar or smaller size compared to the
trajectories of the droplets containing $30\%$ solids in Figure \ref{fig:trajects}.
At the higher solids level in Figure \ref{fig:trajects}, most of
the restructuring observed at both aspect ratios moves the doublets
down the energy gradient, but does not move them into the calculated
energy minimum state, likely because of a larger elastic resistance
to deformation. Interestingly, we observe two instances at $40\%$
solids where the droplets move up the energy gradient, possibly because
of heterogeneities in the solid structure of the droplet. Even more
extreme effects of the larger aspect ratios are observed at the lower
solids concentration studied. 

\begin{figure*}
\includegraphics[scale=0.6]{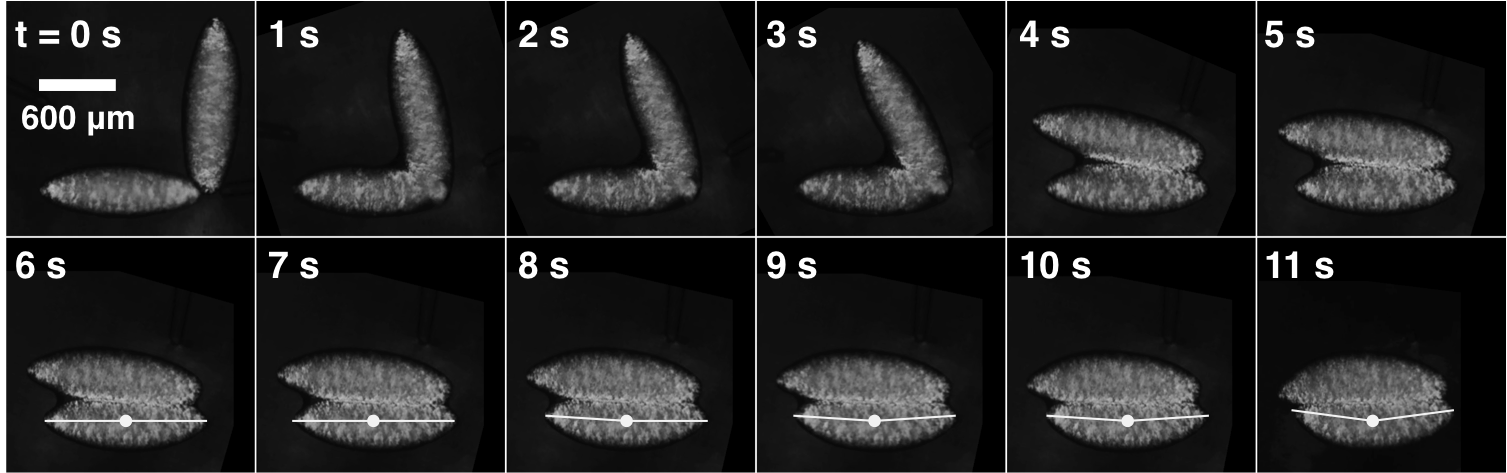}

\caption{\label{fig:highAR} Arrested coalescence of high aspect ratio ellipsoidal
droplets, $AR=3.7$, containing $30\%$ solids. The doublet exhibits
deformation of the ellipsoid being moved, curving it around the second
droplet. }
\end{figure*}

The top row of Figure \ref{fig:highAR} shows a time sequence of images
of two ellipsoids containing $30\%$ solids, with $AR=3.7$, as they
are brought together, begin coalescence, and then arrest. The images
in Figure \ref{fig:highAR} are captured between crossed polarizers
to highlight the solid phase regions of the structures as they move.
In the first few frames of Figure \ref{fig:highAR}, the droplets
move together steadily because of migration of the fluid meniscus,
reducing doublet surface area and moving it from a relatively high
energy configuration, near the saddle-point in the middle of Figure
\ref{fig:Masterplot}A, toward the low-energy region of the upper-right
quadrant of Figure \ref{fig:Masterplot}A. The time interval is the
same for each frame, but the distance moved from the frame at $t=3\:s$
to the frame at $4\:s$ is about $300\,\mu m$, a significant acceleration.
Much like elliptical gears, we find that ellipsoidal droplets can
convert a relatively constant driving force, expansion of the fluid
meniscus, into a variable speed movement. Such forces also have a
strong effect on the individual droplets.

The bottom row of Figure \ref{fig:highAR} highlights the behavior
of the two droplets after reaching a near-final conformation: edge-to-edge
alignment in a low energy state. An overlaid line, connecting the
lower ellipsoid's endpoints with its midpoint, is used to visualize
the subsequent deformation of the ellipsoid as the meniscus continues
to expand completely across the length of the doublet. Here the deflection
of the lower ellipsoid is $\thicksim120$$\:\mu m$ in the last frame,
about $10\%$ of the ellipsoid length. The significant deformation
of the ellipsoids by the meniscus force only occurs once the droplets
have been brought together at the energetic minimum with respect to
their relative orientation. At this point, no further reduction in
the meniscus area can occur by rotations; however deformation of the
ellipsoids can bring their ends into closer contact, reducing the
overall contact area at the expense of elastic reorganization of the
internal structure. The lower solids level of the ellipsoids means
that this deformation is feasible. Past measurements of the petrolatum
system \cite{Caggioni2014} indicate droplets containing $30\%$ solids
have a yield stress, $\sigma_{y}\thicksim1$$\:Pa$ and an interfacial
tension $\gamma\thicksim10\:mN/m$. The stress needed to deform the
droplet must then be of the same order as the droplet yield stress
\cite{Caggioni2014}. Assuming the Laplace pressure sets the magnitude
of applied stress, we can calculate the necessary maximum radius of
the fluid meniscus, R, to cause such a change from:

\begin{equation}
R=\frac{2\gamma}{\sigma_{y}}\label{eq:stress}
\end{equation}
Using Equation \ref{eq:stress}, we solve to obtain $R=20\:mm$ and,
since the measured meniscus radius is more on the order of $R=150\:\mu m$,
conclude that the meniscus force is more than sufficient to cause
the changes observed in Figure \ref{fig:highAR}. Such deformation
is only observed when droplets have low solids levels and high aspect
ratios, but Equation \ref{eq:stress} indicates it could be avoided
by increasing the elastic resistance to droplet deformation \cite{Caggioni2014,Caggioni2015,bayles2018model}.

\section*{Conclusions}

The processes of self-assembly and restructuring of pairs of ellipsoidal
droplets undergoing arrested coalescence have been studied here using
stable droplets, produced in a millifluidic process, with varying
aspect ratios. The ellipsoidal shape of the droplets enables restructuring
and deformation behavior that is significantly different from that
exhibited by spherical droplets \cite{Dahiya:2017gk}. Ellipsoidal
droplets exhibit more complex trajectories, and a wider range of motion,
than spheres \cite{Dahiya:2017gk} during meniscus-driven restructuring.
The experimentally observed dynamics approach, but don't always attain,
theoretical predictions of the most stable configurations with respect
to geometry and surface energy. Small amounts of deformation occur
for all droplets, reducing the radius of their path when compared
to that predicted by the locus of an ellipse rotating around a second
ellipse. The final configuration of the droplets also deviates from
more developed three-dimensional simulations of arrested coalescence
of ellipsoids, likely because of elastic and frictional resistance
to the stresses transmitted by the fluid meniscus. The resistance
can be tuned by adjusting the dimensions and solids fraction of the
droplets, potentially allowing control of larger structure assembly
and restructuring.

We are particularly interested in how irregularly-shaped, non-Brownian,
particles and droplets can be assembled and organized by the action
of a fluid meniscus to achieve changes in shape and to form larger-scale
structures. This work demonstrates limits on the ability to control
shape change and other dynamics using droplets containing solid particles
or viscoelastic microstructures. Varying the aspect ratio of the droplets
controls the extent of restructuring that is possible before a stable
configuration is reached. All ellipsoidal droplets experience some
restructuring, mostly moving toward lower surface energy states. Low
aspect ratio ellipsoids form mostly metastable configurations during
restructuring, while higher aspect ratios mostly converge to energy
minima, though increased elasticity can offset such effects. Tuning
the rheology of the droplets impacts the final configurations, likely
by affecting the significance of local and variable deformation of
individual droplets. The importance of this work is its droplet-level
study of arrested coalescence and restructuring mechanisms for non-spherical
shapes. It is hoped the insights can be used to create accurate predictions
of larger-scale aggregates formed via assembly and meniscus-driven
restructuring \cite{Prileszky:2016gc}, enabling design of emulsion
microstructures and their mechanical properties \cite{bayles2018model}.
Applications are possible in areas as diverse as food product development
\cite{Kim2013}, advanced material creation \cite{feng2013specificity},
and additive manufacturing \cite{Prileszky:2016gc}.

\acknowledgement

This material is based upon work supported by the National Science Foundation under Grant No. DMR-CMMT-1654283.

\providecommand{\latin}[1]{#1}
\makeatletter
\providecommand{\doi}
  {\begingroup\let\do\@makeother\dospecials
  \catcode`\{=1 \catcode`\}=2 \doi@aux}
\providecommand{\doi@aux}[1]{\endgroup\texttt{#1}}
\makeatother
\providecommand*\mcitethebibliography{\thebibliography}
\csname @ifundefined\endcsname{endmcitethebibliography}
  {\let\endmcitethebibliography\endthebibliography}{}

\end{document}